# Multiple Depot Ring Star Problem: A Polyhedral Study and an Exact Algorithm

**Kaarthik Sundar · Sivakumar Rathinam**

**Abstract** The Multiple Depot Ring-Star Problem (MDRSP) is an important combinatorial optimization problem that arises in the context of optical fiber network design, and in applications pertaining to collecting data using stationary sensing devices and autonomous vehicles. Given the locations of a set of customers and a set of depots, the goal is to (i) find a set of simple cycles such that each cycle (ring) passes through a subset of customers and exactly one depot, (ii) assign each non-visited customer to a visited customer or a depot, and (iii) minimize the sum of the routing costs, *i.e.*, the cost of the cycles and the assignment costs. We present a mixed integer linear programming formulation for the MDRSP and propose valid inequalities to strengthen the linear programming relaxation. Furthermore, we present a polyhedral analysis and derive facet-inducing results for the MDRSP. All these results are then used to develop a branch-and-cut algorithm to obtain optimal solutions to the MDRSP. The performance of the branch-and-cut algorithm is evaluated through extensive computational experiments on several classes of test instances.

## 1 Introduction

The Multiple Depot Ring-Star Problem (MDRSP) is an important combinatorial optimization problem arising in the context of optical fiber network design [4,12], and in applications pertaining to collecting data using stationary sensing devices and autonomous vehicles [21]. Given the locations of a set of customers and a set of depots,

K. Sundar *(corresponding author)*,
Graduate student,
Dept. of Mechanical Engineering, Texas A&M University, College Station, TX 77843, USA
E-mail: kaarthiksundar@tamu.edu

S. Rathinam
Assistant Professor,
Dept. of Mechanical Engineering, Texas A&M University, College Station, TX 77843, USA
E-mail: srathinam@tamu.edu

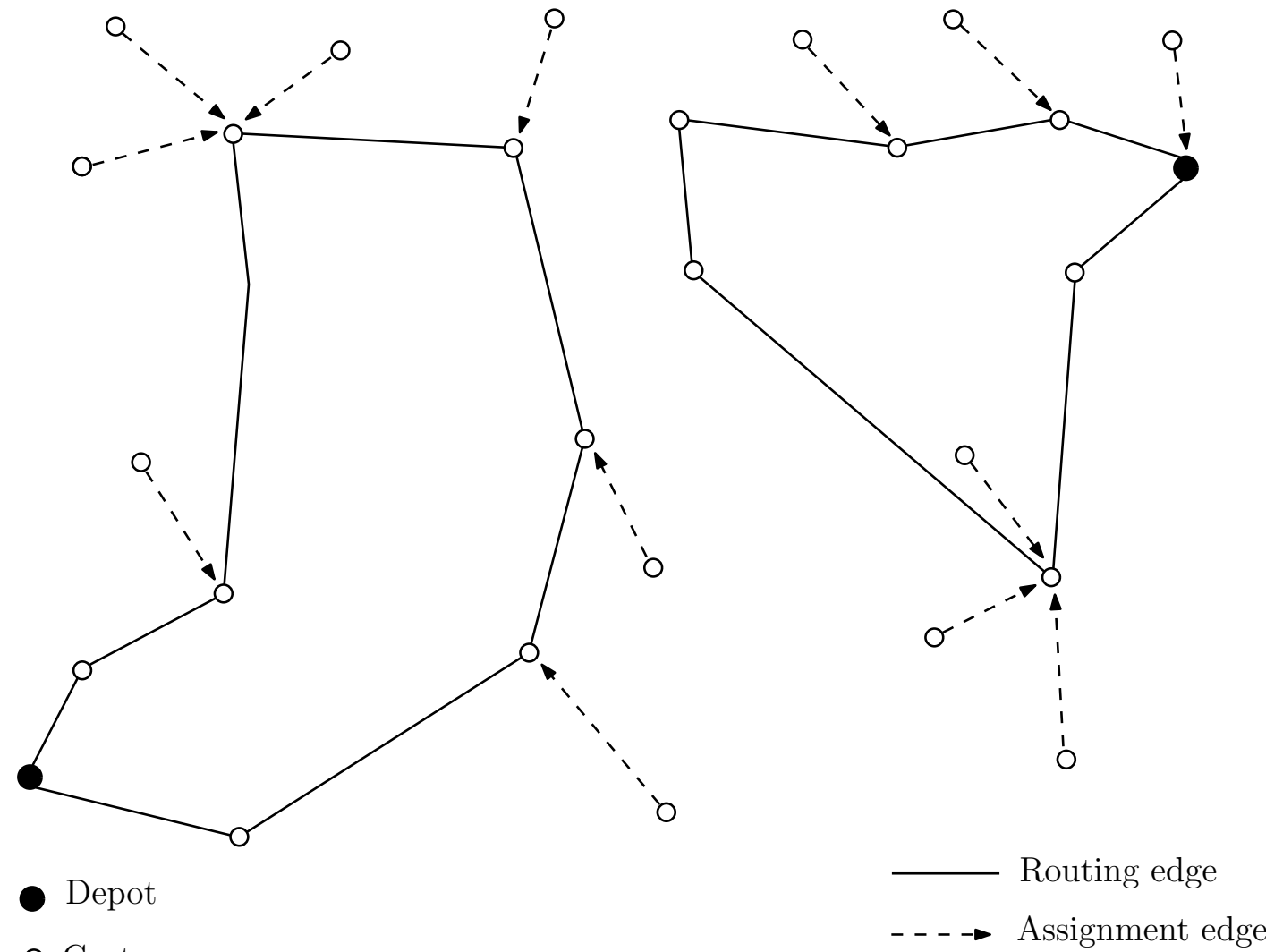

**Fig. 1** Example of a feasible MDRSP solution

the goal is to (i) find a set of simple cycles such that each cycle (ring) passes through a subset of customers and exactly one depot, (ii) assign each non-visited customer to a visited customer or a depot, and (iii) minimize the sum of the routing costs, *i.e.*, the cost of the cycles and the assignment costs. Fig. 1 shows an example of a feasible solution to the MDRSP.

The MDRSP consists of two underlying sub-problems, namely the Traveling Salesman Problem (TSP) and the assignment problem. The two sub-problems are coupled by the fact that the subset of customers that are present in each cycle is not known a priori. If the assignment costs are very large compared to the routing costs, the MDRSP reduces to the multiple depot TSP (MDTSP) [7] and is $\mathcal{NP}$-hard. The single depot variant of the MDRSP, the *ring-star problem* (RSP), has been well studied in the literature. The RSP was first introduced in the context of communication networks in [17] where the authors develop variable neighborhood tabu search algorithms to find good feasible solutions. In [12], the authors present a polyhedral analysis and branch-and-cut algorithms for computing optimal solutions to the RSP. Authors in [13] consider a related problem called the *median-cycle problem* that consists of finding a simple cycle that minimizes the routing cost subject to an upper bound on the total assignment cost. Authors in [13] propose integer linear programming models, introduce additional valid inequalities and implement the model in a branch-and-cut framework.

Several authors have also considered graph structures (other than a cycle) such as a path or a tree [14]. Authors in [16] address a related single-depot problem called the *Steiner ring-star problem*; it consists of finding a minimum cost ring-star in the presence of Steiner vertices. This problem arises frequently in the context of digital data service network design where the objective is to connect terminals to concentrators using point-to-point links (star topology) and to interconnect the concentrators through a ring structure. The authors develop a branch-and-cut algorithm to solve the problem to optimality. A tabu search algorithm was also proposed for the *Steiner ring-star problem* in [23].

The capacitated version of the RSP is also well studied in the literature. Heuristics and exact algorithms based on a branch-and-cut approach are available for a *capacitated multiple ring-star problem* [5]. Heuristics and lower bounds were presented for a capacitated variant of the MDRSP in [4]. A branch-and-cut algorithm to solve the capacitated variant of the MDRSP to optimality was presented in [11]. The authors in [11] also develop a meta-heuristic to obtain feasible solutions. The computational results in [11] indicate that their meta-heuristic outperforms the heuristic proposed by authors in [4] for most of the instances considered.

In this article, we present a mixed integer linear programming formulation for the MDRSP and propose valid inequalities to strengthen the linear programming relaxation. We then present a polyhedral analysis and derive facet-inducing cuts for the MDRSP. Our theoretical results generalize the polyhedral analysis available for the MDTSP [7] and the RSP [12] to the MDRSP. The strengthened formulation is then used to develop a branch-and-cut algorithm to obtain optimal solutions to the MDRSP. The performance of the branch-and-cut algorithm is evaluated through extensive computational experiments on several classes of test instances.

## 2 Problem Description

Let $D := \{r_1, r_2, \ldots, r_n\}$ denote the set of depots. Let $T$ represent the set of customers. Consider a mixed graph $G = (V, E \cup A)$ where $V = D \cup T$, $E$ is a set of undirected edges joining any two distinct vertices in $V$, and $A$ is a set of directed arcs that includes self-loops (*i.e.,* $A = \{[i,j] : i, j \in V\}$). Edges in $E$ refer to the undirected routing edges, and the arcs in $A$ refer to the directed assignment edges. For each edge $(i, j) = e \in E$, we associate a non-negative routing cost $c_e = c_{ij}$, and for each arc $[i, j] \in A$, we associate a non-negative assignment cost $d_{ij}$. Given a subset $E' \subset E$, let $\mathcal{V}(E')$ denote the set of vertices incident to at least one edge in $E'$. Note that we allow the degenerate case where a cycle can only consist of depot and a customer. A ring-star $R$ is denoted by $(V, E' \cup A')$ where $E' \subset E$ is a simple cycle (ring) containing exactly one depot from $D$, and $A' \subseteq A$ is a set of assignment edges (star) between a subset of $T \setminus \mathcal{V}(E')$ and the vertices of $\mathcal{V}(E')$. We say that a customer $i$ is assigned to a ring-star $R$ if it is either visited by the simple cycle (*i.e.,* $i \in \mathcal{V}(E')$) or it is connected to a vertex present in a cycle using an assignment edge (*i.e.,* a vertex $j$ exists such that $[i, j] \in A'$). The cost of the ring star $R$ is the sum of the routing cost of edges in $E'$ and the communication cost of the arcs in $A'$. The objective of the MDRSP is to design at most $n$ ring-stars so that each customer is assigned to exactly one ring-star, and the sum of the cost of all the ring-stars is a minimum.

## 3 Mathematical Formulation

This section presents a mathematical formulation for the MDRSP inspired by the models for the standard routing problems in [7, 12, 22].

We propose a two-index formulation for the MDRSP. We associate to each feasible solution $\mathcal{F}$, a vector $\mathbf{x} \in \mathbb{R}^{|E|}$ (a real vector indexed by the elements of $E$) such that the value of the component $x_e$ associated with edge $e$ is the number of times $e$ appears in the feasible solution $\mathcal{F}$. Note that for some edges $e \in E$, $x_e \in \{0,1,2\}$. If $e$ connects two vertices $i$ and $j$, then $(i,j)$ and $e$ will be used interchangeably to denote the same edge. Similarly, associated to $\mathcal{F}$, is also a vector $\mathbf{y} \in \mathbb{R}^{|A|}$, *i.e.,* a real vector indexed by the elements of $A$. The value of the component $y_{ij}$ associated with a directed arc $[i,j] \in A$ is equal to one if the customer $i$ is assigned to customer $j$ and zero otherwise. Furthermore, we require that a customer $i$ be present in a cycle if it is assigned to itself, *i.e.,* $y_{ii} = 1$.

For any $S \subset V$, we define $\gamma(S) = \{(i,j) \in E : i,j \in S\}$ and $\delta(S) = \{(i,j) \in E : i \in S, j \notin S\}$. If $S = \{i\}$, we simply write $\delta(i)$ instead of $\delta(\{i\})$. Finally, for any $\hat{E} \subseteq E$, we define $x(\hat{E}) = \sum_{(i,j)\in\hat{E}} x_{ij}$, and for any disjoint subsets $A, B \subseteq V$, $x(A:B) = \sum_{i\in A, j\in B} x_{ij}$. Using the above notations, the MDRSP is formulated as a mixed integer linear program as follows:

$$\text{Minimize} \quad \sum_{e\in E} c_e x_e + \sum_{[i,j]\in A} d_{ij} y_{ij} \tag{1}$$

subject to

$$x(\delta(i)) = 2y_{ii} \quad \forall i \in T, \tag{2}$$

$$\sum_{j\in V} y_{ij} = 1 \quad \forall i \in T, \tag{3}$$

$$x(\delta(S)) \geq 2\sum_{j\in S} y_{ij} \quad \forall S \subseteq T, i \in S, \tag{4}$$

$$x(D':\{j\}) + 3x_{jk} + x(\{k\}: D\setminus D') \leq 2(y_{jj} + y_{kk}) \quad \forall j,k \in T; D' \subset D, \tag{5}$$

$$x(D':\{j\}) + 2x(\gamma(S\cup\{j,k\}) + x(\{k\}: D\setminus D') \leq \sum_{v\in S\cup\{j,k\}} 2\, y_{vv} - \sum_{b\in S} y_{ab}$$
$$\forall a \in S; j,k \in T, j \neq k; S \subseteq T\setminus\{j,k\}, S \neq \emptyset; D' \subset D, \tag{6}$$

$$y_{ii} = 1 \quad \forall i \in D, \tag{7}$$

$$y_{ij} = 0 \quad \forall i \in D; j \in T, \tag{8}$$

$$x_{ij} \in \{0,1\} \quad \forall (i,j) \in E; i,j \in T, \tag{9}$$

$$x_{ij} \in \{0,1,2\} \quad \forall (i,j) \in E; i \in D; j \in T, \tag{10}$$

$$y_{ij} \in \{0,1\} \quad \forall [i,j] \in A. \tag{11}$$

In the above formulation, the constraints in (2) ensure the number of undirected (routing) edges incident on any vertex $i \in T$ is equal to 2 if and only if target $i$ is assigned to itself ($y_{ii} = 1$). The constraints in (3) enforce the condition that a vertex $i \in T$ is either in a cycle ($y_{ii} = 1$) or assigned to a vertex $j$ in a cycle (*i.e.,* $y_{ij} = 1$ for some $j \in V$, $j \neq i$). The constraints in (4) are the connectivity or sub-tour elimination constraints. They ensure a feasible solution has no sub-tours of any subset of customers in $T$. The constraints in (5) and (6) are the path elimination constraints. They do not allow for any cycle in a feasible solution to consist of more than one depot. The validity of these constraints is discussed in the subsection 3.1. Constraints in Eq. (7) and (8) are the assignment constraints for the depots. Finally, the constraints (9)-(11) are the integrality restrictions on the $\mathbf{x}$ and $\mathbf{y}$ vectors.

3.1 Path elimination constraints:

The first version of the path elimination constraints was developed in the context of location routing problems [15]. Laporte et al. named these constraints as chain-barring constraints. Authors in [6] and [7] use similar path elimination constraints for the location routing and the multiple depot traveling salesman problems. Any path that originates from a depot and visits exactly two customers before terminating at another depot is removed by the constraint in (5). The validity of the constraint (5) can be easily verified as in [15]. Any other path $d_1, t_1, \cdots, t_p, d_2$, where $d_1, d_2 \in D$, $t_1, \cdots, t_p \in T$ and $p \geq 3$, violates inequality (6) with $D' = \{d_1\}$, $S = \{t_2, \cdots, t_{p-1}\}$, $j = t_1$, $k = t_p$ and $i = t_r$ where $2 \leq r \leq p-1$. We now state and prove a result concerning inequality (4) that will aid in the verifying the validity of the constraint in Eq. (6).

**Lemma 1** *The connectivity constraints in Eq.* (4) *is equivalent to* $x(\gamma(S)) \leq \sum_{v \in S} y_{vv} - \sum_{j \in S} y_{ij}$ *for all* $i \in S$, $S \subseteq T$.

*Proof* Consider a set $S$ with $\emptyset \neq S \subseteq T$. Then, for any feasible solution to the MDRSP, we have the following equality,

$$\sum_{v \in S} x(\delta(v)) = 2x(\gamma(S)) + x(\delta(S))$$

$$\sum_{v \in S} 2y_{vv} \geq 2x(\gamma(S)) + 2\sum_{j \in S} y_{ij} \quad \forall i \in S \text{ (from Eq.4)}$$

$$x(\gamma(S)) \leq \sum_{v \in S} y_{vv} - \sum_{j \in S} y_{ij} \quad \forall i \in S \text{ (from Eq.2)} \tag{12}$$

Hence proved. □

The above lemma states that inequalities (4) and (12) are equivalent and any feasible solution to the MDRSP satisfies both these constraints. We use this equivalence to prove the validity of (6) for the MDRSP. Using the lemma 1, we reduce (6) to

$$x(D' : \{j\}) + 2\,x(\{j\} : S) + 2\,x(\{k\} : S) + x(\{k\} : D \setminus D') + 2x_{jk} \leq 2\,(y_{jj} + y_{kk}) + \sum_{b \in S} y_{ab} + \left( x(\delta(S)) - 2\sum_{b \in S} y_{ab} \right). \tag{13}$$

In any feasible solution to the MDRSP, we will have either $x(\delta(S)) = 2\sum_{b \in S} y_{ab}$ or $x(\delta(S)) > 2\sum_{b \in S} y_{ab}$, since the sub-tour elimination constraint in Eq. (4) is satisfied by all feasible solutions. First, we claim that no feasible solution satisfying $x(\delta(S)) = 2\sum_{b \in S} y_{ab}$ is removed by (13). Consider a feasible solution $\mathcal{F}$ that satisfies $x(\delta(S)) = 2\sum_{j \in S} y_{ij}$. $\mathcal{F}$ can either have $\sum_{j \in S} y_{ij} = 1$ or $\sum_{j \in S} y_{ij} = 0$. When $\sum_{j \in S} y_{ij} = 0$, inequality (13) reduces to

$$x(D' : \{j\}) + x(\{k\} : D \setminus D') + 2x_{jk} \leq 2\,(y_{jj} + y_{kk})$$

which is trivially satisfied by the corresponding feasible solution. When $\sum_{j \in S} y_{ij} = 1$, we obtain

$$x(D' : \{j\}) + 2x(\{j\} : S) + 2x(\{k\} : S) + x(\{k\} : D \setminus D') + 2x_{jk} \leq 2\,(y_{jj} + y_{kk}) + 1. \tag{14}$$

The above inequality is also satisfied by the feasible solution $\mathcal{F}$. (refer to Appendix for the proof).

A similar proof can also be given to the case when a feasible solution satisfies $x(\delta(S)) > 2\sum_{j\in S} y_{ij}$ (refer to the appendix for the proof). Therefore, the path elimination constraints in Eq. (6) are valid to the MDRSP.

We note that our formulation allows for a feasible solution with paths connecting two depots and visiting exactly one customer. In the literature, such paths are referred to as 2-paths. As the formulation allows for two copies of an edge between a depot and a target, 2-paths can be eliminated and therefore, there always exists an optimal solution which does not contain any 2-path. In the following subsection, we shall strengthen the linear programming relaxation of the model (2)-(11) by the introduction of additional valid inequalities.

### 3.2 Additional Valid Inequalities

In this section, we develop three classes of valid inequalities for the MDRSP. Consider the constraints in Eq. (4). For any $S = \{i, j\}$ where $i, j \in T$, Eq.(4) reduces to $x(\delta(i)) + x(\delta(j)) - 2x_{ij} \geq 2y_{ii} + 2y_{ij}$ . Further simplification using Eq. (2) yields

$$x_{ij} \leq y_{jj} - y_{ij}. \tag{15}$$

As in [12], one can note that the constraints in (15) dominate the following traditional inequalities:

$$x_{ij} \leq 1 - y_{ij},\ x_{ij} \leq y_{ii} \text{ and } x_{ij} \leq y_{jj} \quad \text{for all } i, j \in T.$$

Another set of useful constraints similar to (15) is given by

$$x_{ij} \leq 2y_{jj} \quad \text{for all } i \in D, j \in T. \tag{16}$$

Inequalities valid for a TSP polytope are also valid for the MDRSP. We particularly examine the *2-matching inequalities* available for the TSP polytope [9]. Specifically, we consider the following inequality:

$$x(\gamma(H)) + x(\mathcal{T}) \leq \sum_{i\in H} y_{ii} + \frac{|\mathcal{T}| - 1}{2} \tag{17}$$

for all $H \subseteq T$ and $\mathcal{T} \subset \delta(H)$. Here $H$ is called the handle, and $\mathcal{T}$ the teeth. $H$ and $\mathcal{T}$ satisfy the following conditions:

- the edges in the teeth are not incident to any depots in the set D,
- no two edges in the teeth are incident on the same customer,
- $|\mathcal{T}| \geq 3$ and odd.

The 2-matching inequality is also valid for the RSP [12]; the proof of validity of the inequality for the MDRSP is given in following proposition.

**Proposition 1** *The 2-matching inequality in Eq.* (17) *is valid for any feasible solution to the MDRSP.*

*Proof* For any $H \subseteq T$ and $\mathcal{T} \subset \delta(H)$ satisfying the conditions stated previously, we have the following equality:

$$
\begin{aligned}
2x(\gamma(H)) + x(\delta(H)) &= \sum_{v \in H} x(\delta(v)) \\
\Rightarrow\ 2x(\gamma(H)) + x(\mathcal{T}) + x(\delta(H) \setminus \mathcal{T}) &= 2 \sum_{v \in H} y_{vv} \quad \text{(from Eq. (2))}.
\end{aligned}
$$

We also have $x(\mathcal{T}) \leq |\mathcal{T}|$ for the set $\mathcal{T}$ (since $x_e \leq 1$ for any $e = (i,j), i, j \in T$). Adding this inequality to the above equality, we obtain,

$$
\begin{aligned}
2x(\gamma(H)) + 2x(\mathcal{T}) + x(\delta(H) \setminus \mathcal{T}) &\leq 2 \sum_{v \in H} y_{vv} + |\mathcal{T}| \\
\Rightarrow\ 2x(\gamma(H)) + 2x(\mathcal{T}) &\leq 2 \sum_{v \in H} y_{vv} + |\mathcal{T}|.
\end{aligned}
$$

The last inequality follows because $x(\delta(H) \setminus \mathcal{T}) \geq 0$. Further simplification yields

$$
\begin{aligned}
&2x(\gamma(H)) + 2x(\mathcal{T}) \leq 2 \sum_{v \in H} y_{vv} + (|\mathcal{T}| - 1) \quad \text{(since L.H.S is even and } |\mathcal{T}| \text{is odd)} \\
&\Rightarrow x(\gamma(H)) + x(\mathcal{T}) \leq \sum_{v \in H} y_{vv} + \frac{|\mathcal{T}| - 1}{2}.
\end{aligned}
$$

Hence the 2-matching inequality is valid for the MDRSP. □

The constraints in Eq. (17) are also equivalent to the *blossom's inequality* for the 2-matching problem and a special case of the *comb inequality* for the symmetric TSP [3]. Eq. (17) is a comb inequality where the cardinality of every tooth is two and both the handle and the teeth contain only vertices from set $T$.

The next set of valid inequalities is derived using the valid inequalities for the *Stable Set* polytope (SSP). In any feasible solution to the MDRSP, for any triplet of vertices $i, j, k \in T$, the assignments $y_{ij}$ and $y_{ik}$ are incompatible when $j \neq k$. The stable set problem associated with these incompatible assignments is a relaxation of the MDRSP polytope. A similar observation was made for the RSP in [12]. This property leads to the following odd-hole inequalities for the MDRSP:

$$
y_{ij} + y_{jk} + y_{ki} \leq 1 \quad \text{for all } i, j, k \in T \text{ and } i \neq j,\ j \neq k,\ i \neq k. \tag{18}
$$

$$
x(\delta(S)) \geq 2(y_{ij} + y_{jk} + y_{ki}) \quad \text{for all } i, j, k \in T,\ i \neq j,\ j \neq k,\ i \neq k \text{ and } S \subseteq T \text{ such that } i, j, k \in S. \tag{19}
$$

Eq. (19) is the valid inequality obtained from the two previously mentioned relaxations of the MDRSP, *i.e.,* the SSP and TSP relaxations. In the following section, we develop some polyhedral results and facet-inducing properties for the valid inequalities discussed thus far.

## 4 Polyhedral Analysis

We will show the polyhedral results for the MDRSP while leveraging on the results already known for a multiple depot TSP (MDTSP). MDTSP is a special case of the MDRSP when each customer must be visited by one of the vehicles. Let $P$ denote the polytope that represents the convex hull of feasible solutions to the MDRSP (*i.e.,* satisfies (2)-(11)) and $Q$ denote the corresponding MDTSP polytope [7].

If $u$ denotes the number of customers, we observe that there are $u$ equalities in (2), $u$ equalities in (3), $n$ equalities in (7) and $nu$ equalities in (8). Therefore, the system (2), (3), (7) and (8) has $2u+n+nu$ equalities. We also note that this system of equality constraints are linearly independent.

The number of $x_e$ variables in the formulation is $\binom{u}{2} + nu$ ($\binom{u}{2}$ is the number of edges between customers and $nu$ is the number of edges between depots and customers). Similarly, the number of $y_{ij}$ variables in the formulation is $u^2 + n + 2nu$ ($u^2$ is the number of customer to customer arcs, $n$ is the number of arcs that assigns a depot to itself and $2nu$ is the number of arcs that assigns a depot to a customer and vice versa). Let $m$ denote the total number of variables used in the problem formulation *i.e.,* $m = \binom{u}{2} + u^2 + n + 3nu$.

Let $\chi_{(\mathbf{x},\mathbf{y})} \in \mathbb{R}^m$ denote the incidence vector of a solution $(\mathbf{x}, \mathbf{y})$ to the MDRSP in the graph $G$. Now we have,

$$P := \text{conv}\{\chi_{(\mathbf{x},\mathbf{y})} : (\mathbf{x}, \mathbf{y}) \text{ is a feasible MDRSP solution}\} \tag{20}$$

$$Q := \{(\mathbf{x}, \mathbf{y}) \in P : y_{ii} = 1 \text{ for all } i \in T\} \tag{21}$$

The dimension of the polytope $Q$ was shown to be $\binom{u}{2} + u(n-1)$ in [7]. Let $F \subseteq T$ denote a subset of customers. To relate $P$ and $Q$, we define an intermediate polytope $P(F)$ as follows:

$$P(F) := \{(\mathbf{x}, \mathbf{y}) \in P : y_{ii} = 1 \text{ for all } i \notin F\}. \tag{22}$$

We observe that, $P(\emptyset) = Q$ and $P(T) = P$. Also, for any $(\alpha, \beta) \in \mathbb{R}^m$ and $\gamma \in \mathbb{R}$, define the hyperplane

$$\mathcal{H}(\alpha, \beta, \gamma) := \{(\mathbf{x}, \mathbf{y}) \in \mathbb{R}^m : \alpha\mathbf{x} + \beta\mathbf{y} = \gamma\} \tag{23}$$

**Lemma 2** *Let $v_1, \cdots, v_u$ be an ordering of the customers in the set $T$ and $F_k = \{v_1, \cdots, v_k\}$ for all $k \in \{1, \cdots, u\}$. If for each $k = 1, \ldots, u$ and each $v_l \in V \setminus \{v_k\}$, there exists a feasible solution to the MDRSP, such that*

1. *$y_{v_j v_j} = 1$ for all $j > k$ i.e., every customer in the set $T \setminus F_k$ is in some cycle,*
2. *$y_{v_j v_j} + \sum_{r \in D} y_{v_j r} = 1$ for all $j < k$ i.e., every vertex in the set $F_k$ must be either in a cycle or assigned to a depot,*
3. *$y_{v_k v_l = 1}$ i.e., the vertex $v_k$ must be assigned to the vertex $v_l$, and*
4. *$\alpha\mathbf{x} + \beta\mathbf{y} = \gamma$,*

*then,* $\dim(P \cap \mathcal{H}(\alpha, \beta, \gamma)) \geq \dim(Q \cap \mathcal{H}(\alpha, \beta, \gamma)) + u(u + n - 1)$.

*Proof* We prove by induction on $|F_k|$ that $\dim(P(F_k) \cap \mathcal{H}(\alpha, \beta, \gamma)) \geq \dim(Q \cap \mathcal{H}(\alpha, \beta, \gamma)) + |F_k|(u + n - 1)$. This in turn proves the lemma because when $F_k = T$, we have $P(F_k) = P$ and $|F_k| = u$. Let $N_k := \dim(Q \cap \mathcal{H}(\alpha, \beta, \gamma)) + |F_k|(u+n-1)$. The base case for induction holds since $k = 0$ implies $F_k = \emptyset$ and $P(F_k) = Q$. Now, suppose $k > 0$. Then by induction hypothesis, we have $\dim(P(F_{k-1}) \cap \mathcal{H}(\alpha, \beta, \gamma)) \geq$ $N_{k-1}$. Hence, there are at least $N_{k-1}+1$ affine independent points in the polytope $P(F_{k-1}) \cap \mathcal{H}(\alpha, \beta, \gamma)$. All these affine independent points satisfy $y_{v_k v_k} = 1$ (since $v_k \notin F_{k-1}$). From the definition of $P(F)$ in Eq. (22), we have $P(F_k) \cap \mathcal{H}(\alpha, \beta, \gamma) \supset P(F_{k-1}) \cap \mathcal{H}(\alpha, \beta, \gamma)$. Therefore, these $N_{k-1}+1$ affine independent points satisfying $y_{v_k v_k} = 1$ (Fig. 2) lie in $P(F_k) \cap \mathcal{H}(\alpha, \beta, \gamma)$. The assumptions of the lemma provide for additional $(u+n-1)$ affine independent points in $P(F_k) \cap \mathcal{H}(\alpha, \beta, \gamma)$ that satisfy $y_{v_k v_k} = 0$ (Fig. 2). Therefore, $\dim(P(F_k) \cap \mathcal{H}(\alpha, \beta, \gamma)) \geq N_{k-1} + (u+n-1) = \dim(Q \cap \mathcal{H}(\alpha, \beta, \gamma)) + |F_k|(u+n-1)$. Hence proved. □

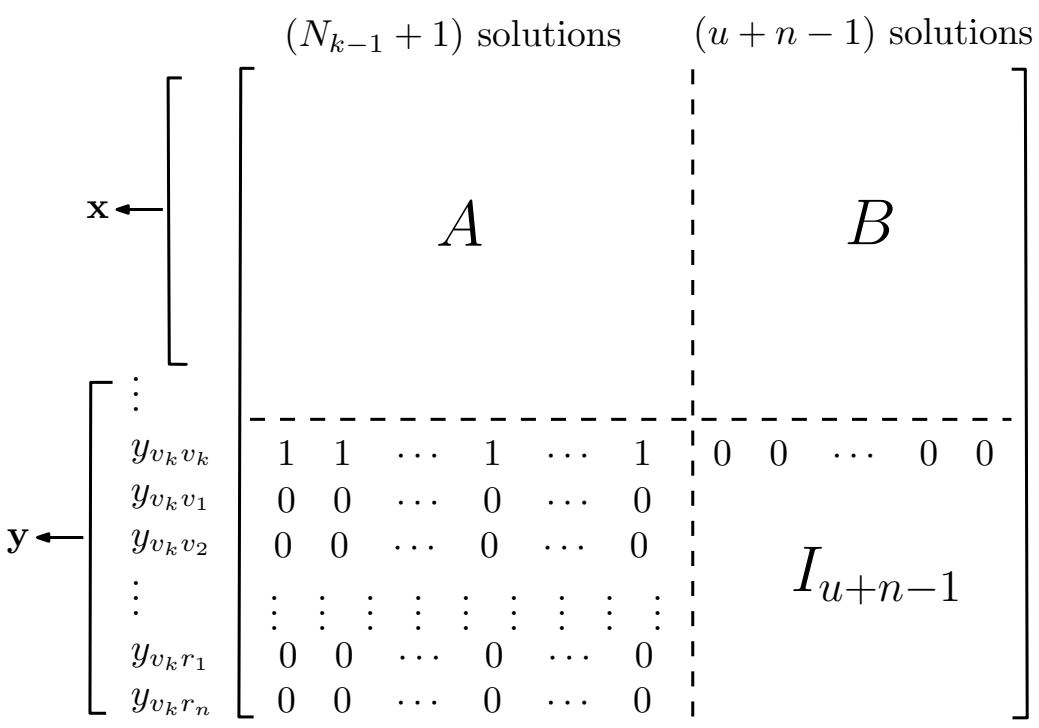

**Fig. 2** The figure shows the affine independent feasible solutions. The induction hypothesis provides the affine independence of the $N_{k-1}+1$ solutions in $P(F_k) \cap \mathcal{H}(\alpha, \beta, \gamma)$. Additional $(u+n-1)$ affine independent solutions are provided by the assumptions in the lemma. All the solutions put together are affine independent due to the block diagonal structure of the solution matrix.

The Lemma 2's hypothesis provides a family of feasible solutions to the MDRSP that is guaranteed to be linearly independent. The dimension of the MDRSP polytope $P$ is computed in the following corollary of Lemma 2.

**Corollary 1** $\dim(P) = \binom{u}{2} + u^2 + 2u(n-1)$.

*Proof* The number of variables used in formulation of MDRSP is $\binom{u}{2} + u^2 + n + 3nu$ and all the solutions of the MDRSP satisfy the $2u + n + nu$ linearly independent equality constraints in the system (2, 3, 7, 8). Hence, $\dim(P) \leq \binom{u}{2} + u^2 + n + 3nu - (2u + n + nu) = \binom{u}{2} + u^2 + 2u(n-1)$. Also, we have,

$$
\begin{aligned}
\dim(P) &= \dim(P \cap \mathcal{H}(0,0,0)) \\
&\geq \dim(Q \cap \mathcal{H}(0,0,0)) + u(u+n-1) \qquad \text{(using Lemma 2)} \\
&= \dim(Q) + u^2 + u(n-1) \\
&= \binom{u}{2} + u(n-1) + u^2 + u(n-1) \text{ (see Benavent and Martínez [2013])} \\
&= \binom{u}{2} + u^2 + 2u(n-1)
\end{aligned}
$$

Hence, $\dim(P) = \binom{u}{2} + u^2 + 2u(n-1)$. □

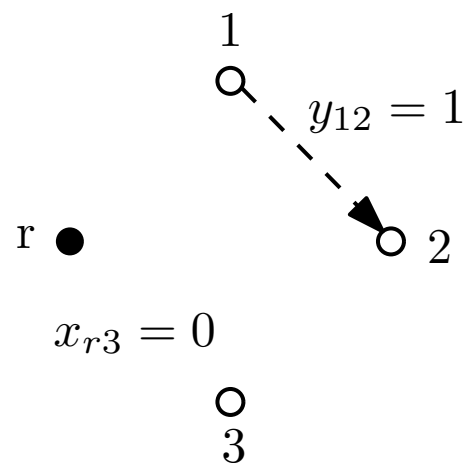

**Fig. 3** When $|T| < 4$ in Prop. (2), a feasible solution to the MDRSP with customers 2,3 in the cycle associated with depot $r$ such that $x_{r3} = 0$ and $y_{12} = 1$ cannot be constructed.

An important consequence of Lemma 2 is that any valid inequality $\alpha\mathbf{x} + \beta\mathbf{y} \leq \gamma$ that is facet-inducing to the MDTSP polytope $Q$ and satisfying the conditions (1)–(4) of the lemma is valid and facet-inducing to the MDRSP polytope $P$. This observation will be used in all of the subsequent results concerning the polyhedral analysis of $P$.

**Proposition 2** *If $|T| \geq 4$, the inequality $x_e \geq 0$ is facet-inducing for $P$ for every $e \in E$.*

*Proof* For any ordering of the customers in $T$, it is trivial to construct feasible solutions satisfying the conditions *1–4* of Lemma 2 ($x_e = 0$ is the hyperplane here) for a fixed $e = (i, j) \in E$. To construct such feasible solutions satisfying the assumptions of the Lemma, we require the condition $|T| \geq 4$ (refer to Fig. 3). The proposition follows by noting that $x_e = 0$ is a facet to the MDTSP polytope $Q$ if $|T| \geq 4$ (see [7]). □

*Remark 1* We also note that the inequality $x_{ij} \leq 1$ for all $(i, j) \in E$ and $i, j \in T$ is not facet-inducing for $P$ since it is dominated by the constraint in Eq. (15). Similarly, the inequality $x_{ij} \leq 2$ for all $(i, j) \in E$, $i \in D$ and $j \in T$ is not facet-inducing for the polytope $P$ as it is dominated by the corresponding constraint in Eq. (16).

**Proposition 3** *The sub-tour elimination constraint given by Eq.* (4)*, i.e., $x(\delta(S)) \geq 2\sum_{j\in S} y_{ij}$ is facet-inducing for the MDRSP polytope for each $S \subseteq T$, $i \in S$, $|S| \geq 2$.*

*Proof* Consider any ordering of the customers in set $T$ such that the customer $i \in T$ is in the last position of the ordering. We will prove the proposition by constructing feasible solutions satisfying assumptions of Lemma 2 ( $x(\delta(S)) = 2\sum_{j\in S} y_{ij}$ is the hyperplane here) for the considered ordering.
Choose an arbitrary customer $k \in T \setminus \{i\}$. Given $k$, we construct $(u + n - 1)$ feasible solutions satisfying the assumptions of the Lemma 2 as follows: construct a cycle spanning all the customers in $T \setminus \{k\}$ and some depot $r$ with exactly 2 edges in $\delta(S)$ and customer $k$ assigned to any vertex in the set $V \setminus \{k\}$ (illustrated in Fig. 4-(a)). The cardinality of the set $V \setminus \{k\}$ is $(u + n - 1)$ and hence we obtain $(u + n - 1)$ feasible solutions satisfying the assumptions of the Lemma.
We now detail the procedure for constructing another $(u + n - 1)$ feasible solutions for the last customer $i \in T$. Construct a cycle spanning depot $r$ and all the customers in $S \setminus \{i\}$ with exactly two edges in $\delta(S)$ while assigning $i$ to any vertex in $S \setminus \{i\}$. This provides for $|S| - 1$ feasible solutions that satisfy the assumptions of the Lemma 2. Another set of $|V \setminus S|$ feasible solutions is obtained as follows: construct a cycle spanning the depot $r$ and the vertex set $T \setminus S$, and assign the customers in $S \setminus \{i\}$ to one of the depots and the customer $i$ to any vertex in the set $V \setminus S$ (illustrated in Fig. 4-(b)). This final set of feasible solutions ensure $x(\delta(S)) = 0$ and $2\sum_{j\in S} y_{ij} = 0$. The proposition then follows because $x(\delta(S)) \geq 2\sum_{j\in S} y_{ij}$ reduces to a facet-inducing inequality $x(\delta(S)) \geq 2$ for the polytope $Q$ of the MDTSP (see [7]). □

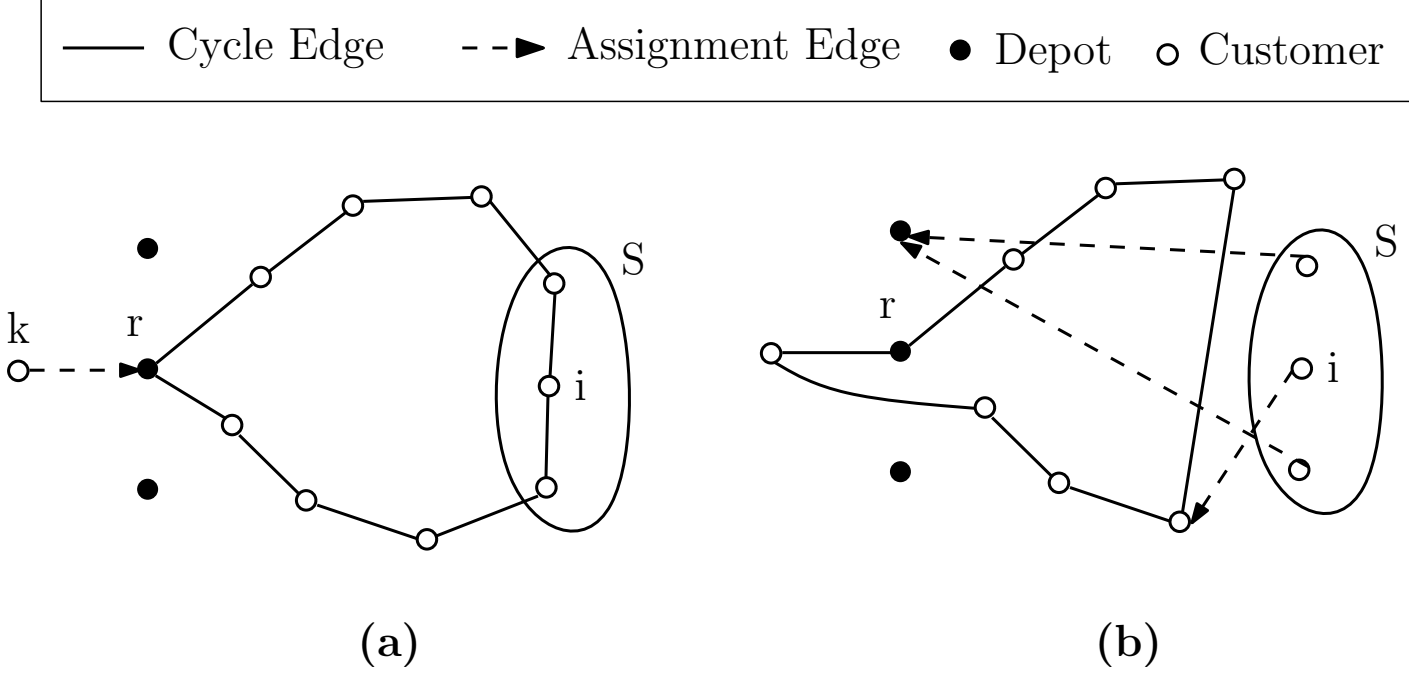

**Fig. 4** Feasible solutions described in Prop. (3). (a) shows a feasible solution where customer $k$ is assigned to the depot $r$. We generate $(u+n-1)$ solutions be changing the vertex to which $k$ is assigned. (b) shows a feasible solution where the customer $i$ is assigned to a vertex in the set $V \setminus S$, with a cycle spanning the customers in $V \setminus S$. Changing the assignment for customer $i$ would result in $|V \setminus S|$ feasible solutions.

*Remark 2* The Prop. 3 does not hold for $|S| = 1$, since the degree constraint in Eq. (2) dominates the corresponding constraint with $|S| = 1$. Similarly, when $i \notin S$, authors in [12] showed that Prop. 3 is not valid for the RSP because of the inequality

$$x(\delta(S \cup \{i\})) = x(\delta(S)) + x(\delta(i)) - 2\sum_{j\in S} x_{ij} \geq 2\sum_{j\in S\cup i} y_{ij} = 2(y_{ii} + \sum_{j\in S} y_{ij}).$$

The above inequality implies $x(\delta(S)) \geq 2\sum_{j\in S}(y_{ij} + x_{ij})$ which dominates the corresponding constraint in Eq. (4) when $i \notin S$. The same argument holds for the MDRSP.

**Proposition 4** *The constraint given by Eq.* (5), $x(D' : \{j\}) + 3x_{jk} + x(\{k\} : D \setminus D') \leq 2(y_{jj} + y_{kk})$*, is facet-inducing for the MDRSP polytope $P$ for every $j, k \in T$, $j \neq k$, $|T| \geq 2$, $D' \subset D$, and $D' \neq \emptyset$.*

*Proof* We shall again use Lemma 2 to prove the proposition. Given $j, k \in T$ and $D' \subset D$, consider any ordering of the vertices in $T$ where $j$ and $k$ appear in the last two positions. We also assume $r_1 \in D'$ and $r_2 \in D \setminus D'$. We claim that there exists a feasible solution for every vertex $i \in T$ and for each vertex $v \in V \setminus \{i\}$ that satisfy the assumptions *1–3* of Lemma 2 and the equation $x(D' : \{j\}) + 3x_{jk} + x(\{k\} : D \setminus D') = 2(y_{jj} + y_{kk})$. This claim combined with the known result that $x(D' : \{j\}) + 3x_{jk} + x(\{k\} : D \setminus D') \leq 4$ is facet-inducing for the MDTSP polytope $Q$ (see [7]) proves the proposition. We shall now prove our claim.

For any arbitrary customer $i \in T \setminus \{j, k\}$, consider the following solutions to the MDRSP: a cycle spanning the depot $r_1$ and all the customers in $T \setminus \{i\}$ such that the customer $j$ is adjacent to the depot $r_1$ and customer $k$ with the customer $i$ assigned to any vertex in the set $V \setminus \{i\}$. Each of these solutions is feasible to the MDRSP and satisfy the equation $x(D' : \{j\}) + 3x_{jk} + x(\{k\} : D \setminus D') = 2(y_{jj} + y_{kk}) = 4$ (since $x(D' : \{j\}) = 1$ and $x_{jk} = 1$). For the customer $j$, consider feasible solutions where $j$ is assigned to a vertex in $V \setminus \{j\}$, the vertex $k$ is the lone vertex spanned the cycle associated with depot $r_2$ and all the customers in $T \setminus \{j, k\}$ are spanned by cycle associated with depot $r_1$. These solutions satisfy the equation $x(D' : \{j\}) + 3x_{jk} + x(\{k\} : D \setminus D') = 2(y_{jj} + y_{kk}) = 2$ (since $x(\{k\} : D \setminus D') = 2$). A similar construction can also be done for the vertex $k$. Therefore the claim, and as a result, the proposition is true. □

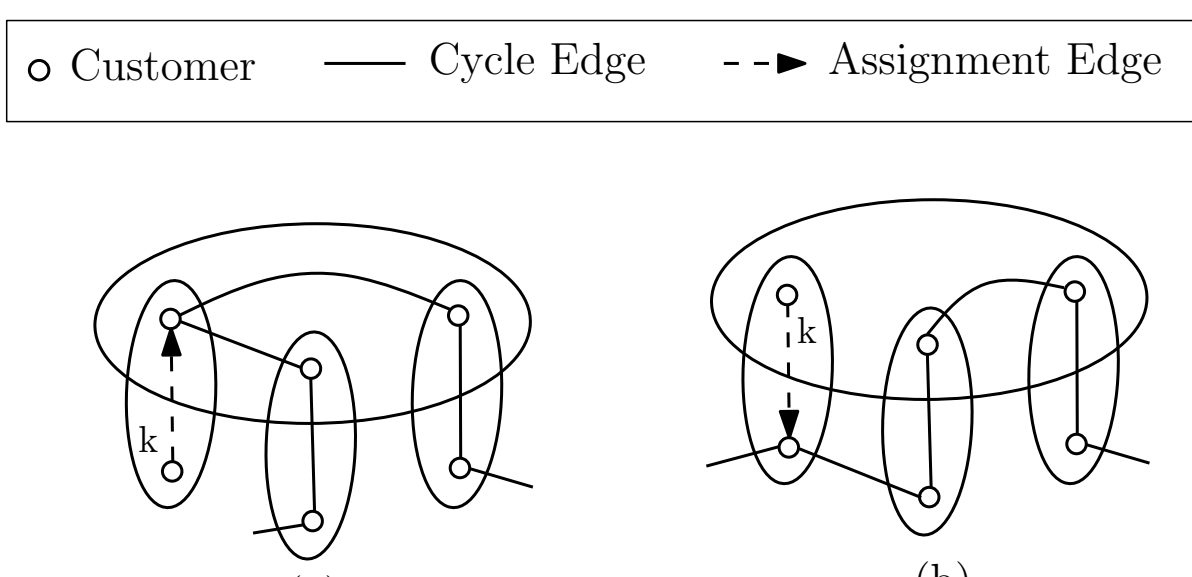

**Fig. 5** Feasible solutions described in Prop. (5). The figures consider a handle and teeth with with $|H| = 3$ and $|\mathcal{T}| = 3$. (a) shows the part of a cycle that spans the vertices in the handle and the teeth when $k \in T$ is in a teeth such that the *2-matching* inequality is satisfied at equality. (b) is a similar figure when $k \in T$ is in the handle $H$.

**Proposition 5** *The 2-matching inequality in Eq.* (17) *for all $H \subseteq T$ and $\mathcal{T} \subset \delta(H)$, satisfying the conditions:*

1. *the edges in the teeth are not incident to any depots in the set $D$,*
2. *no two edges in the teeth are incident on the same customer,*
3. *$|\mathcal{T}| \geq 3$ and odd.*

*is facet-inducing for the MDRSP polytope $P$ when $|T| \geq 6$.*

*Proof* The proof proceeds by constructing feasible solutions that satisfy conditions *1–3* of the Lemma 2 and the hyperplane $x(\gamma(H)) + x(\mathcal{T}) = \sum_{i \in H} y_{ii} + \frac{|\mathcal{T}|-1}{2}$. For a fixed $H$ and $\mathcal{T}$ satisfying the conditions stated in the proposition, and for each $k \in T$, it is straightforward to construct a cycle spanning some depot $r \in D$ and all the customers in $T \setminus \{k\}$ such that $x(\gamma(H)) + x(\mathcal{T}) = \sum_{i \in H} y_{ii} + \frac{|\mathcal{T}|-1}{2}$ (refer to Fig. 5). Each of these cycles can be converted to a feasible solution by the addition of an assignment from customer $k$ to a vertex in the set $V \setminus \{k\}$. The figures show portions of the cycle when $k \in T$ is in the handle and teeth respectively. In Fig. 5–(a), the vertex $k$ is in the handle $H$ and in Fig. 5–(b), $k$ is in a tooth. We also note that the valid inequality $x(\gamma(H)) + x(\mathcal{T}) \leq \sum_{i \in H} y_{ii} + \frac{|\mathcal{T}|-1}{2}$ reduces to $x(\gamma(H)) + x(\mathcal{T}) \leq |H| + \frac{|\mathcal{T}|-1}{2}$ for a MDTSP. The proposition follows since the hyperplane defined by $x(\gamma(H)) + x(\mathcal{T}) \leq |H| + \frac{|\mathcal{T}|-1}{2}$ is a facet for the MDTSP polytope $Q$ when $|T| \geq 6$ (see [7]). □

## 5 Separation Algorithms

In this section, we discuss the algorithms that are used to find violated families of constraints described in Sec. 3. We denote by $G^* = (V^*, E^*)$ the *support graph* associated with a given fractional solution $(\mathbf{x}^*, \mathbf{y}^*)$ *i.e.,* $V^* := \{i \in V : y^*_{ii} > 0\}$ and $E^* := \{(i,j) \in E : x^*_{ij} > 0\}$. We also define $A^* := \{[i,j] \in A : y^*_{ij} > 0\}$.

### 5.1 Separation of Sub-tour Elimination constraints in Eq. (4) and Eq. (15):

As shown previously, the inequalities in Eq. (4) reduce to Eq. (15) when $|S| = 2$. The violation of the inequality in Eq. (15) can be verified by examining the inequality for every pair of customers in the set $T$. Next, we examine the connected components in $G^*$. Each connected component $C$ such that $D \cap C = \emptyset$ generates a violated sub-tour elimination constraint for $S = C$ and for each $i \in S$. If a connected component $C$ has $D \cap C \neq \emptyset$, the following procedure is used to find the largest violated sub-tour elimination constraint in $x(\delta(S)) \geq 2\sum_{j\in S} y_{ij}$ . For any $S \subseteq T$, given any $i \in S$, we can rewrite the constraint in Eq. (4) as

$$x(\delta(S)) + 2\sum_{j \notin S} y_{ij} \geq 2 \qquad \forall S \subseteq T, i \in S. \tag{24}$$

Given a connected component $C$ such that $D \cap C \neq \emptyset$, $i \in C \cap T$, and a fractional solution $(\mathbf{x}^*, \mathbf{y}^*)$, the most violated constraint (24) can be obtained by computing a minimum $s$-$t$ cut on a capacitated undirected graph $\bar{G} = (\bar{V}, \bar{E})$, with $\bar{V} = (V^* \cap T) \cup \{s\}$. The vertex $s$ denotes the source vertex and is formed by contracting all the depots into a single vertex. The vertex $t$ denotes the sink vertex and $t = i$. The edge set $\bar{E} = E^* \cup \{(s,j) : j \in V^* \cap T\}$. Every edge $(s,j)$ where $j \in (V^* \cap T) \setminus \{i\}$ is assigned a capacity $\sum_{d\in D} x^*_{dj}$. The edge $(i,j)$ where $j \in \bar{V} \setminus \{i\}$ is assigned a capacity equal to $x^*_{ij} + 2y^*_{ij}$ and any remaining edge $e$ is assigned a capacity $x^*_e$. We now compute the minimum $s$-$t$ cut $(S, \bar{V} \setminus S)$ with $t \in \bar{V} \setminus S$. The vertex set $S' = \bar{V} \setminus S$ defines the most violated inequality if the capacity of the cut is strictly less than two. A similar separation procedure is also used to separate the sub-tour elimination constraints in [12,5].

### 5.2 Separation of Path-Elimination constraints - Eq. (5) and Eq. (6):

We first discuss the procedure used to separate violated constraints in Eq. (5). Consider every pair of targets $j, k \in T \cap V^*$. We rewrite the constraint in (5) as $x(D' : \{j\}) + x(\{k\} : D \setminus D') \leq 2(y_{kk} + y_{jj}) - 3x_{jk}$. Given $j, k$ and fractional solution $(\mathbf{x}^*, \mathbf{y}^*)$, the RHS of the above inequality is a constant and is equal to $2(y^*_{kk} + y^*_{jj}) - 3x^*_{jk}$. We observe that the LHS of the inequality, $x^*(D' : \{j\}) + x^*(\{k\} : D \setminus D')$, is maximum when $D' = \{d \in D : x^*_{jd} \geq x^*_{kd}\}$. Furthermore, when $D' = \emptyset$ or $D' = D$, no path constraint in Eq.(5) is violated for the given pair of vertices $j$ and $k$. With $D' = \{d \in D : x^*_{jd} \geq x^*_{kd}\}$, if $x^*(D' : \{j\}) + x^*(\{k\} : D \setminus D')$ is strictly greater than $2(y^*_{kk} + y^*_{jj}) - 3x^*_{jk}$, the path constraint in Eq. (5) is violated for the pair of vertices $j, k$ and the subset of depots $D'$.

We now discuss the separation procedure for the the constraint in Eq. (6). We note that this path constraint is determined by a pair of vertices $j, k \in T$, a subset of vertices $S \subseteq T \setminus \{j, k\}$, a vertex $a \in S$ and a subset of depots $D' \subset D$. In what follows we develop a procedure that is applied to every pair of clients $\{j, k\}$. It is obvious that (6) will never be violated if $j$ and $k$ belong to different connected components of the support graph $G^*$; hence, we only consider pairs of those $\{j, k\}$ belonging to the same connected component in $G^*$. We denote $\bar{S} = S \cup \{j,k\}$. Using this notation, we reformulate the constraint in Eq. (6) to Eq. (25), whose violation can be deduced using a minimum $s$-$t$ cut algorithm. The reduction is shown below:

$$x(D' : \{j\}) + x(\{k\} : D \setminus D') + 2x(\gamma(\bar{S})) \leq \sum_{v \in \bar{S}} 2\, y_{vv} - \sum_{b \in S} y_{ab}$$

$$\Rightarrow x(D' : \{j\}) + x(\{k\} : D \setminus D') \leq x(\delta(\bar{S})) - \sum_{b \in S} y_{ab}$$

$$\Rightarrow x(D' : \{j\}) + x(\{k\} : D \setminus D') + 1 \leq x(\delta(\bar{S})) + \sum_{b \notin \bar{S}} y_{ab} + y_{aj} + y_{ak}$$

$$\Rightarrow x(D' : \{j\}) + x(\{k\} : D \setminus D') + 1 - y_{aj} - y_{ak} \leq x(\delta(\bar{S})) + \sum_{b \notin \bar{S}} y_{ab}. \tag{25}$$

The second inequality follows by applying Lemma 1 to the set $\bar{S}$. Eq. (25) is an equivalent representation of the path constraint in Eq. (6). Now, given a fractional solution $(\mathbf{x}^*, \mathbf{y}^*)$, the pair $\{j,k\}$ in the same connected component $C$, and an $a \in (C \setminus \{j,k\}) \cap T$, the LHS of (25) attains a maximum value for $D' = \{d \in D : x^*_{jd} \geq x^*_{kd}\}$ (when $D' = \emptyset$ or $D' = D$, the corresponding path constraint (6) is not violated). Let $\mathcal{L} = x^*(D' : \{j\}) + x^*(\{k\} : D \setminus D') + 1 - y^*_{aj} - y^*_{ak}$. Now, the most violated constraint (6) can be found by computing a minimum $s$-$t$ cut on a capacitated undirected graph $\bar{G} = (\bar{V}, \bar{E})$ with $\bar{V} = V^* \cup \{s,t\}$. The vertex $s$ denotes the source. The vertex $t$ denotes the sink and is formed by contracting all the depots to a single vertex. We add edges with very large capacity from the source vertex $s$ to vertices $j,k$ and $a$. Every edge $(i,a)$ where $i \in V^* \setminus \{a\}$ is assigned a capacity $x^*_{ai} + y^*_{ai}$ and any remaining edge $e$ is assigned a capacity $x^*_e$. The minimum $s$-$t$ cut $(S', T')$ on $\bar{G}$ would have $j,k,a,s \in S'$ and $t,r \in T'$ for every $r \in D$. The pair $j,k$, the vertex set $S = S' \setminus \{s\}$ and the vertex $a \in S$ defines the most violated inequality if the capacity of the cut is strictly less than $\mathcal{L}$.

### 5.3 Separation of 2-Matching constraints - Eq. (17):

We discuss exact and heuristic separation procedures for the 2-matching constraints. Using a construction similar to the one proposed in [19] for the $b$-matching problem, the separation problem for 2-matching inequalities can be transformed into a minimum capacity odd cut problem; hence this separation problem is exactly solvable in polynomial time. This procedure is computationally intensive, and so we use the following simple heuristic proposed by the authors in [8]. We consider an undirected graph $\bar{G} = (\bar{V}, \bar{E})$ with $\bar{V} = V^* \cap T$ and $\bar{E} = \{e : 0 < x^*_e < 1\}$. Then, we consider each connected component $H$ of $\bar{G}$ as a handle of a possibly violated 2-matching inequality whose two-node teeth correspond to the edges $e \in \delta(H)$ with $x^*_e = 1$. We reject the inequality if the number of teeth is even. The procedure can be implemented in $O(|\bar{V}| + |\bar{E}|)$ time.

### 5.4 Separation of constraints in Eq. (18) and Eq. (19):

For the constraints in Eq. (18) and Eq. (19), we use the separation procedures discussed in [12]. The inequalities in Eq. (18) can be separated by a complete enumeration of $i, j, k \in T$ such that $y^*_{ij} > 0$, $y^*_{jk} > 0$ and $y^*_{ki} > 0$. Similarly, for each $i, j, k \in T$ such that $y^*_{ij} > 0$, $y^*_{jk} > 0$ and $y^*_{ki} > 0$, a min-cut separating $D$ from $\{i, j, k\}$ in $G^*$ would detect the most violated constraint in Eq. (19), if any.

## 6 Branch-and-Cut Algorithm

In this section, we describe important implementation details of the branch-and-cut algorithm for the MDRSP. The algorithm is implemented within a CPLEX 12.4 framework using the CPLEX callback functions [1]. The callback functions in CPLEX enable the user to completely customize the branch-and-cut algorithm embedded into CPLEX including, the choice of node to explore in the enumeration tree, the choice of branching variable, the separation and the addition of user-defined cutting planes and the application of heuristic methods.

The lower bound at the root node of the enumeration tree is computed by solving the LP relaxation of the formulation in Sec. 3 that is further strengthened using the cutting planes described in Sec. 3.2. The initial linear program consisted of all constraints in (1)-(11) and (16), except (4), (5) and (6). Several numerical experiments indicated that the inequalities in Eq. (18) and Eq. (19) were not computationally helpful for the branch-and-cut procedure, and hence were not used in the final implementation of the algorithm. For a given LP solution, we identify violated inequalities using the separation procedures in the following order: (i) sub-tour elimination constraints in Eq. (15), (ii) sub-tour elimination constraints in Eq. (4) (iii) path elimination constraints in Eq. (5) and Eq. (6) (iv) 2-matching constraints in Eq. (17). Furthermore, we disabled the separation of all the cuts embedded into the CPLEX framework because enabling these cuts increased the average computation time for the instances. Once the new cuts generated using these separation procedures were added to the linear program, the tighter linear program was resolved. This procedure was iterated until either of the following conditions was satisfied: (i) no violated constraints could be generated by the separation procedures, (ii) the current lower bound of the enumeration tree was greater or equal to the current upper bound. If no constraints are generated in the separation phase, we create subproblems by branching on a fractional variable. First, we select a fractional $y_{ii}$ variable, based on the *strong branching* rule [2]. If all these variables are integer, then we select a fractional $x_e$ variable using the same rule. As for the node-selection rule, we used the best-first policy for all our computations, *i.e.,* select the subproblem with the lowest objective value.

### 6.1 Heuristics

We discuss a greedy algorithm called *LP-heuristic,* that aides in speeding up the convergence of the branch-and-cut algorithm. The *LP-heuristic* constructs a feasible solution from a given fractional LP solution. It is used only at the root node of the enumeration tree, once in every three iterations. *LP-heuristic* is based on a transformation method [18]. Given $\mathbf{y}^*$, the vector of fractional LP assignment values, the

**Procedure -** *Greedy Assignment*
**Input:** $\mathbf{y}^*$;
**Output:** assignments $\sigma$, set $P$ of vertices that are spanned by some cycle;
**comment:** initialization
**for each** $i \in T$ **do** $\sigma(i) := -1$;
$\bar{T} := T$; **comment:** customers to be assigned
$P := V$; **comment:** vertices that are spanned by some cycle
**comment:** customer assignment
**while** $\bar{T} \neq \emptyset$ **do**
  Select a customer $i \in \bar{T}$ randomly; $\bar{T} = \bar{T} \setminus \{i\}$;
  $\sigma(i) = \text{argmax}\{y^*_{ik} : k \in P\}$;
  **if** $\sigma(i) \neq i$ **then** $P = P \setminus \{i\}$;
**endwhile**

**Fig. 6** The greedy assignment procedure

heuristic greedily assigns every customer in the set $T$ to some vertex in the set $V$. We call this procedure the greedy assignment procedure; a pseudo-code of the algorithm is shown in Fig 6. Once we have the assignment, we can compute the set of vertices that are spanned by some cycle (the set of vertices that are assigned to itself). We then solve the multiple depot traveling salesmen problem (MDTSP) on these vertices and $D$. A heuristic based on the transformation method [18] and LKH heuristic [10] is used to solve the MDTSP.

## 7 Computational Results

In this section, we discuss the computational results of the branch-and-cut algorithm. The algorithm was implemented in C++ (GCC version 4.6.3), using the elements of Standard Template Library (STL) and CPLEX 12.4 framework. As mentioned in Sec. 6, the internal CPLEX cut generation was disabled and hence, CPLEX was used only to manage the enumeration tree. All the simulations were performed on a Dell Precision T5500 workstation (Intel Xeon E5630 processor @2.53 GHz, 12 GB RAM). The computation times reported are expressed in seconds and we imposed a time limit of 7200 seconds for each run of the algorithm. The performance of the algorithm was tested on different classes of test instances, all generated using the traveling salesman problem library [20].

*Instance generation:* We generated two classes of test instances (A and B) having the same underlying graph, but with a different assignment cost structure (similar to [5,12]). For each of the two classes and for each value of $|T| \in \{29, 51, 76, 101\}$, we generated 12 MDRSP instances using four TSPLIB instances [20] namely, *bays29, eil51, eil76* and *eil101.* We performed a computational study on these instances with $|D| \in \{3, 4, 5\}$. The depot locations were randomly generated. The routing costs and assignment costs were generated as follows:

*Class I:* The routing and assignment cost for a pair of vertices $i, j$ is equal to the Euclidean distance $l_{ij}$ between the two vertices.

*Class II:* For each pair of vertices $i, j$, the routing cost $c_{ij} = \alpha l_{ij}$ and the assignment cost $d_{ij} = (10 - \alpha)l_{ij}$ where $\alpha \in \{3, 5, 7, 9\}$. We refer to $\alpha$ as the scale factor.

| Name | $\|D\|$ | opt | %-LB | Pair | SEC | 2mat | PEC | Nodes | Time |
|---|---|---|---|---|---|---|---|---|---|
| bays29 | 3 | 8040.9424 | 99.23 | 51 | 676 | 6 | 1675 | 17 | 7.65 |
| bays29 | 4 | 8040.9424 | 99.23 | 49 | 682 | 6 | 1675 | 17 | 7.47 |
| bays29 | 5 | 7651.3744 | 100 | 42 | 423 | 1 | 1218 | 0 | 1.92 |
| eil51 | 3 | 391.6344 | 100 | 75 | 959 | 6 | 0 | 0 | 3.01 |
| eil51 | 4 | 383.7392 | 99.99 | 72 | 1313 | 5 | 102 | 2 | 10.2 |
| eil51 | 5 | 377.7437 | 99.97 | 69 | 1156 | 5 | 1733 | 3 | 18.2 |
| eil76 | 3 | 496.0161 | 99.83 | 131 | 2708 | 19 | 2172 | 40 | 265.03 |
| eil76 | 4 | 494.5647 | 99.74 | 140 | 3196 | 57 | 4243 | 115 | 126.63 |
| eil76 | 5 | 491.6595 | 99.60 | 130 | 3511 | 38 | 5516 | 257 | 265.64 |
| eil101 | 3 | 589.93 | 99.93 | 174 | 3739 | 8 | 1388 | 3 | 210.22 |
| eil101 | 4 | 585.8615 | 99.92 | 177 | 5312 | 5 | 1303 | 5 | 250.61 |
| eil101 | 5 | 581.255 | 99.96 | 170 | 4684 | 4 | 4744 | 3 | 427.45 |
| Averages | | | 99.78 | 106.67 | 2363.25 | 13.33 | 2147.42 | 38.50 | 132.84 |

**Table 1** Computational results for instances in *Class I.*

Tables 1–2 summarize the computational behavior of the branch-and-cut algorithm on the two classes of instances. The column headings are defined as follows:

**Name:** instance name (for Classes I and II);
$|D|$**:** number of depots (for Classes I and II);
$\alpha$**:** scale factor (for Class II);
**opt:** optimal objective value (for Classes I and II);
**%-LB:** percentage LB/opt, where LB is the objective value of the LP relaxation computed at the root node of the enumeration tree (for Classes I and II);
**Pair:** number of constraints (15) generated (for Classes I and II);
**SEC:** number of constraints (4) with $|S| > 2$ generated (for Classes I and II);
**2mat:** number of constraints (17) generated (for Classes I and II);
**PEC:** number of constraints (5) and (6) generated (for Classes I and II);
**Nodes:** total number of nodes examined in the enumeration tree (for Classes I and II);
**Time:** total computation time in seconds (for Classes I and II).

The results indicate that the proposed branch-and-cut algorithm can solve instances involving up to 101 customers with modest computation times. We observe that the Class II instances are more difficult, on an average, especially for a scale factor equal to 5 and 7. The difficult instances tend to be those with relatively few vertices in the cycles. The %-LB column in both the tables indicate that the lower bound obtained at the root node of the enumeration tree is very tight, typically within 0.5% of the optimum. The average number of 2-matching valid inequalities in (17) generated in the enumeration tree is less when compared to the other inequalities. This indicates that the proposed mixed integer linear programming formulation for the MDRSP is by itself very tight. Overall, we were able to solve all the 60 TSPLIB based instances, with the largest instance involving 101 customers and 5 depots.

| Name | \|D\| | $\alpha$ | opt | %-LB | Pair | SEC | 2mat | PEC | Nodes | Time |
|---|---|---|---|---|---|---|---|---|---|---|
| bays29 | 3 | 3 | 27180.6246 | 98.08 | 4 | 309 | 24 | 402 | 54 | 3.89 |
| bays29 | 4 | 3 | 27180.6246 | 98.08 | 4 | 322 | 25 | 416 | 56 | 3.95 |
| bays29 | 5 | 3 | 26389.0317 | 98.31 | 4 | 339 | 34 | 582 | 129 | 4.46 |
| bays29 | 3 | 5 | 40204.7118 | 99.23 | 50 | 643 | 9 | 875 | 37 | 6.39 |
| bays29 | 4 | 5 | 40204.7118 | 99.23 | 48 | 670 | 11 | 912 | 38 | 6.55 |
| bays29 | 5 | 5 | 38256.8719 | 99.93 | 40 | 442 | 1 | 956 | 3 | 2.87 |
| bays29 | 3 | 7 | 41155.8684 | 99.80 | 120 | 313 | 0 | 545 | 3 | 6.44 |
| bays29 | 4 | 7 | 41155.8684 | 99.77 | 120 | 314 | 0 | 635 | 3 | 7.01 |
| bays29 | 5 | 7 | 39812.3467 | 97.29 | 127 | 526 | 6 | 2420 | 379 | 35.19 |
| bays29 | 3 | 9 | 19016.3546 | 100.00 | 250 | 60 | 0 | 0 | 0 | 0.09 |
| bays29 | 4 | 9 | 17279.8835 | 100.00 | 219 | 13 | 0 | 0 | 0 | 0.06 |
| bays29 | 5 | 9 | 15149.4721 | 100.00 | 177 | 13 | 0 | 0 | 0 | 0.05 |
| eil51 | 3 | 3 | 1279.7286 | 98.95 | 7 | 1200 | 92 | 3741 | 250 | 67.92 |
| eil51 | 4 | 3 | 1268.1921 | 98.82 | 7 | 1145 | 65 | 3519 | 323 | 70.82 |
| eil51 | 5 | 3 | 1266.1108 | 98.69 | 11 | 1432 | 157 | 8831 | 1266 | 409.76 |
| eil51 | 3 | 5 | 1958.1716 | 100.00 | 75 | 959 | 6 | 0 | 0 | 6.12 |
| eil51 | 4 | 5 | 1918.6961 | 99.99 | 72 | 1313 | 5 | 30 | 2 | 21.4 |
| eil51 | 5 | 5 | 1888.7183 | 99.97 | 72 | 1422 | 5 | 516 | 10 | 43.53 |
| eil51 | 3 | 7 | 1969.3083 | 99.87 | 233 | 704 | 0 | 1208 | 3 | 106.75 |
| eil51 | 4 | 7 | 1822.9502 | 97.94 | 265 | 4166 | 8 | 23102 | 7688 | 4935.53 |
| eil51 | 5 | 7 | 1789.38 | 97.49 | 268 | 4481 | 50 | 27655 | 12297 | 5485.37 |
| eil51 | 3 | 9 | 806.9033 | 100.00 | 398 | 20 | 0 | 0 | 0 | 0.2 |
| eil51 | 4 | 9 | 722.8627 | 100.00 | 352 | 20 | 0 | 0 | 0 | 0.2 |
| eil51 | 5 | 9 | 710.4597 | 100.00 | 348 | 20 | 0 | 0 | 0 | 0.21 |
| eil76 | 3 | 3 | 1636.4959 | 99.87 | 8 | 499 | 13 | 310 | 3 | 24.72 |
| eil76 | 4 | 3 | 1636.4959 | 99.66 | 8 | 512 | 20 | 412 | 15 | 17.99 |
| eil76 | 5 | 3 | 1633.4431 | 99.65 | 8 | 615 | 21 | 753 | 29 | 21.41 |
| eil76 | 3 | 5 | 2480.0803 | 99.83 | 134 | 4545 | 43 | 3774 | 114 | 326.11 |
| eil76 | 4 | 5 | 2472.8233 | 99.74 | 135 | 3382 | 20 | 3480 | 71 | 93.07 |
| eil76 | 5 | 5 | 2458.2974 | 99.60 | 131 | 3859 | 37 | 5414 | 255 | 286.04 |
| eil76 | 3 | 7 | 2468.3815 | 99.44 | 376 | 1553 | 1 | 2290 | 13 | 160.97 |
| eil76 | 4 | 7 | 2424.515 | 99.25 | 360 | 1364 | 4 | 2455 | 11 | 193.55 |
| eil76 | 5 | 7 | 2370.4893 | 98.65 | 390 | 4633 | 101 | 11196 | 1770 | 1386.5 |
| eil76 | 3 | 9 | 1455.2214 | 99.96 | 876 | 1070 | 0 | 1541 | 3 | 244.9 |
| eil76 | 4 | 9 | 1427.1416 | 100.00 | 822 | 1082 | 0 | 1125 | 0 | 26.59 |
| eil76 | 5 | 9 | 1184.0794 | 100.00 | 703 | 451 | 0 | 232 | 0 | 4.65 |
| eil101 | 3 | 3 | 1918.0064 | 99.49 | 8 | 1246 | 68 | 1778 | 95 | 185.16 |
| eil101 | 4 | 3 | 1903.5985 | 99.36 | 15 | 1596 | 69 | 1586 | 157 | 133.66 |
| eil101 | 5 | 3 | 1899.0916 | 99.56 | 11 | 883 | 24 | 720 | 47 | 311.75 |
| eil101 | 3 | 5 | 2949.65 | 99.93 | 174 | 3739 | 8 | 1388 | 3 | 220.56 |
| eil101 | 4 | 5 | 2929.3075 | 99.92 | 177 | 5312 | 5 | 1303 | 5 | 256.89 |
| eil101 | 5 | 5 | 2906.2748 | 99.96 | 170 | 4684 | 4 | 4744 | 3 | 443.73 |
| eil101 | 3 | 7 | 2895.725 | 99.57 | 488 | 2073 | 0 | 2558 | 9 | 338.19 |
| eil101 | 4 | 7 | 2842.6639 | 99.86 | 484 | 1662 | 0 | 2773 | 5 | 249.39 |
| eil101 | 5 | 7 | 2828.3891 | 99.45 | 543 | 4627 | 0 | 22379 | 2899 | 5523.73 |
| eil101 | 3 | 9 | 1805.7752 | 99.45 | 1362 | 1956 | 0 | 82 | 36 | 134.23 |
| eil101 | 4 | 9 | 1601.3474 | 99.19 | 1166 | 1422 | 0 | 405 | 120 | 154.41 |
| eil101 | 5 | 9 | 1565.0956 | 99.32 | 1126 | 1230 | 0 | 141 | 48 | 119.64 |
| Averages: | | 3 | | 99.04 | 7.92 | 841.50 | 51.00 | 1920.83 | 202.00 | 104.62 |
| | | 5 | | 99.78 | 106.50 | 2580.83 | 12.83 | 1949.33 | 45.08 | 142.77 |
| | | 7 | | 99.03 | 314.50 | 2201.33 | 14.17 | 8268.00 | 2090.00 | 1535.72 |
| | | 9 | | 99.83 | 649.92 | 613.08 | 0.00 | 293.83 | 17.25 | 57.10 |

**Table 2** Computational results for instances in Class II

## 8 Conclusion

In this paper, we have presented an exact algorithm for the MDRSP, a problem that arises in the context of design of optical fiber network in telecommunications, resource allocation in monitoring applications, to name a few. A mixed integer linear programming formulation including several classes of valid inequalities was proposed and a complete polyhedral analysis with facet-inducing results were investigated together with a branch-and-cut algorithm. The algorithm was tested on a wide class of benchmark instances from a standard library. The largest solved instance involved 101 vertices. Future work can be directed towards development of branch-and-cut approaches accompanied with a polyhedral study to solve capacitated versions of the problem.

## A Appendix

**Proposition 6** *Let $j, k, a \in T$, $S \subseteq T \setminus \{j, k\}$, $a \in S$, $D' \subset D$ such that $D'$ and $S$ are non-empty. Consider any feasible solution $\mathcal{F}$ to the MDRSP, that satisfies $\sum_{b \in S} y_{ab} = 1$ and $x(\delta(S)) = 2\sum_{b \in S} y_{ab}$. Then $\mathcal{F}$ satisfies the constraint $x(D' : \{j\}) + 2x(\{j\} : S) + 2x(\{k\} : S) + x(\{k\} : D \setminus D') + 2x_{jk} \leq 2(y_{jj} + y_{kk}) + 1$.*

*Proof* Let us first consider the case where $y_{jj} = 0$. Then by the degree constraints in Eq. 2, we have $x(D' : \{j\}) = x(\{j\} : S) = x_{jk} = 0$. Hence the constraint reduces to $2x(\{k\} : S) + x(\{k\} : D \setminus D') \leq 2y_{kk} + 1$, which is trivially satisfied by $\mathcal{F}$. A similar argument holds for the case where $y_{kk} = 0$. Now, consider the case where $y_{jj} = y_{kk} = 1$. Then the RHS of the constraint takes the value 5. It is not difficult to observe that for any feasible solution that satisfies the assumptions of the proposition, the maximum value the LHS of the constraint is 5.

□

**Proposition 7** *Let $j, k, a \in T$, $S \subseteq T \setminus \{j, k\}$, $a \in S$, $D' \subset D$ such that $D'$ and $S$ are non-empty. Consider any feasible solution $\mathcal{F}$ to the MDRSP, that satisfies $x(\delta(S)) > 2\sum_{b \in S} y_{ab}$. Then $\mathcal{F}$ satisfies the constraint $x(D' : \{j\}) + 2\,x(\{j\} : S) + 2\,x(\{k\} : S) + x(\{k\} : D \setminus D') + 2x_{jk} \leq 2(y_{jj} + y_{kk}) + \sum_{b \in S} y_{ab} + \left(x(\delta(S)) - 2\sum_{b \in S} y_{ab}\right)$.*

*Proof* Similar to the proof of the last proposition, let us first consider the case where $y_{jj} = 0$. Then the constraint reduces to $2x(\{k\} : S) + x(\{k\} : D \setminus D') \leq 2y_{kk} + \sum_{b \in S} y_{ab} + (x(\delta(S)) - 2\sum_{b \in S} y_{ab})$. When $y_{kk} = 0$, this constraint is trivially satisfied. So, we assume $y_{kk} = 1$. Then, $\mathcal{F}$ has $x(\delta(S)) > 2\sum_{b \in S} y_{ab}$. The minimum value of $x(\delta(S)) - 2\sum_{b \in S} y_{ab}$ is 2 (from lemma (1)). Hence the minimum value the RHS of the constraint is 4 and it is also observed that the maximum value the LHS is 4. Hence $\mathcal{F}$ satisfies the constraint when $y_{jj} = 0$. A similar holds for the case where $y_{kk} = 0$. Now consider the case where $y_{jj} = y_{kk} = 1$. It can also be observed that, the minimum value taken by the RHS and the maximum value taken by the LHS of the constraint is 6 and hence the constraint is satisfied by $\mathcal{F}$.

□